\documentclass[review]{elsarticle}
\usepackage{setspace,graphicx,subfigure,caption,color,amsmath,amssymb,setspace,lineno,hyperref,graphics}
\usepackage{epstopdf}

\usepackage{footmisc}
\journal{Pervasive and Mobile Computing}
\makeatletter
\newcommand\footnoteref[1]{\protected@xdef\@thefnmark{\ref{#1}}\@footnotemark}
\makeatother
\newcommand{\RR}[1]{\textcolor{black}{#1}}
\newcommand{\ca}[1]{\textcolor{black}{#1}}
\newcommand{\DA}[1]{\textcolor{black}{#1}}
\begin{document}

\begin{frontmatter}

\title{Continuous Gait Velocity Estimation using Houseohld Motion Detectors}
\author[mymainaddress]{Rajib Rana\corref{mycorrespondingauthor}}
\ead{to.rajib.rana@gmail.com}

\author[address3]{Daniel Austin}
\ead{austidan@ohsu.edu}

\author[address4]{Peter~G.~Jacob}
\ead{jacobsp@ohsu.edu}

\author[mysecondaryaddress]{Mohanraj~Karunanithi}
\cortext[mycorrespondingauthor]{Corresponding author}
\ead{mohan.Karunanithi@csiro.au}

\author[address3]{Jeffrey Kaye}
\ead{kaye@ohsu.edu}

\address[mymainaddress]{Autonomous Systems Laboratory, CSIRO, Australia}
\address[mysecondaryaddress]{Australian E-Health Research Centre, CSIRO, Australia}

\address[address3]{Department of Neurology, Oregon Health \& Science University, Portland, OR 97239 USA}

\address[address4]{Departments of Biomedical Engineering and Otolaryngology, Oregon Health and Science University, Portland, OR 97239 USA.}

\begin{abstract}
%\RR{Daniel, here we want to push that sensor-line gait velocity estimation is great way of doing things. However, it suffers measurement sparsity for two reasons. First, peole may not go though all the sensors in the array. Second, due to higher installation sensor-array cannot be installed in multiple locations at home. This not only cause measurement sparsity but also provides gait velocity in only one fixed location. {\bf Third, due to higher maintenance cost, sensor-array is not sustainable for long-term gait velocity estimation.} However, continuous and location-specific gait velocity measurement is of clinical significance. We propose an alternative approach, where we deploy and run only one sensor-array for a short period of time to collect base-line gait velocity information. We then develop a model based approach which using the base-line measurement as ground truth , predicts the \emph{location-specific} gait velocity from transition time.}
Gait velocity has been consistently shown to be an important indicator and predictor of health status, especially in older adults. Gait velocity is often assessed clinically, but the assessments occur infrequently and thus do not allow optimal detection of key health changes when they occur.  In this paper, we show the time it takes a person to move between rooms in their home - denoted ``transition times" - can predict gait velocity when estimated from passive infrared motion detectors installed in a patient's own home.  Using a support vector regression approach to model the relationship between transition times and gait velocities, we show that velocity can be predicted with an average error  less than $2.5$ cm/sec.  This is demonstrated with data collected over a 5 year period from $74$ older adults monitored in their own homes. This method is simple and cost effective, and has advantages over competing approaches such as: obtaining 20-100x more gait velocity measurements per day, and \RR{offering the fusion} of location-specific information with time stamped gait estimates.  These advantages allow stable estimates of gait parameters (maximum or average speed, variability) at shorter time scales than current approaches.  This also provides a pervasive in-home method for context aware gait velocity sensing that allows for monitoring of gait trajectories in space and time.

\end{abstract}


\end{frontmatter}

%\linenumbers

\section{Introduction}
%\RR{It would be better if we could separate out the literature review? We also have to describe here what we mean by location-specific gait velocity. It might be a good idea to use Fig.1 to describe the approach. Daniel could you please glaze up the intro?}
%\DA{Rajib, I looked at about 5 PMC articles and it seems that about half seperate the intro and lit review, and the other half leave them integrated.  I have a slight preference for leaving them integrated so did not split them apart here but feel free to do so if you prefer it that way.}
Gait velocity, also referred to as the speed of walking, is both an important \ca{indicator of individuals' current health} state and predictor of future adverse cognitive and physical health outcomes.  Gait velocity can distinguish between patients with dementia and healthy controls~\cite{bruce2012relationship} and has been shown to decrease prior to cognitive impairment~\cite{buracchio2010trajectory} and in Alzheimer's disease~\cite{Goldman1999}.  Decreased gait velocity is prevalent in dementia~\cite{Beauchet2008} and is predictive of future hospitalization~\cite{Studenski2003}.  Gait is known to require substantial cognitive resources~\cite{Faulkner2006} and gait velocity may be directly related to several cognitive processes such as attention and executive function~\cite{Holtzer2006,Holtzer2012}.  Gait has also been linked to risk of falls~\cite{Kelsey2012,Cuaya2013} and risk of future disability~\cite{Guralnik2000a,studenski2011gait}.

Gait velocity is most commonly assessed clinically with a stopwatch timed walk - such as the 25-ft timed walk~\cite{Larson2013} - although more comprehensive assessments are also used.  A large shortcoming of clinic-based assessments is infrequent test administration.  Often, 6 months to a year or more passes between assessments~\cite{hagler2010unobtrusive}, making it difficult to detect acute changes when they occur or to distinguish between abrupt changes in function and slower changes occurring over time.  Many pervasive computing approaches have been successfully proposed and validated to estimate gait velocity that overcome the infrequency problem of the current clinical assessment methodologies. The existing approaches can be grouped into two categories.  

The first category is based on instrumenting the body with a worn device, such as an accelerometer~\cite{Culhane2005,Gietzelt2013,Dalton2013}.  Accelerometry is accurate and effective but is not well suited for studies lasting more than a few days without substantial requirements of the patients or research/clinical staff participation.  In particular, patients must remember to wear the device, 
\ca{position it correctly on the body, regularly charge the device, and follow any procedures needed for ensuring download and transmission of the data.}
%place it in the correct place on the body, charge the device overnight, and download the data before the storage on the device is filled.  
This is especially problematic in older and cognitively impaired populations who may benefit the most from long term gait monitoring, as they may forget or be unable to perform the tasks required for obtaining reliable and continuous data.  Research staff can mitigate some of these shortcomings at the expense of increased cost, however this drastically reduces scalability for big studies requiring large-scale device deployment.  Accelerometry is also not location-aware (unless additional sensors are added to the system) and thus gait velocity estimates obtained from accelerometry cannot immediately be associated with the activity performed during the walking event (e.g., whether someone is headed toward the bathroom). %\ca{The method presented in this paper does not require a body-worn sensor to measure walking speed.  The system we present also enables \emph{location-aware} velocity estimations, which is not possible even with body-worn sensors.}

\ca{Approaches to device-free velocity tracking are typically} based on instrumenting the environment (most often the home) with unobtrusive sensors.  Strategies in this category include the use of ``restricted" infrared sensors arranged in a walking line~\cite{hagler2010unobtrusive,kaye2012one}, referred to as a ``sensor-line'' or \ca{a camera-based sensors such as a Microsoft Kinect system} to estimate gait~\cite{Clark2013,Stone2013,sivapalan2011compressive}.  Both the camera and sensor-line methods overcome some of the issues posed by the body worn devices and are readily deployable for very long term monitoring and in large scale studies.  However, However, these methods have limitations in capturing unobtrusive monitoring of gait.  For example, both systems only detect walks when a resident passes within the field-of-view of the sensors.  This can result in data sparsity when a subject passes through the instrumented area infrequently. The higher installation and maintenance cost (as compared to the method presented below) and dedicated equipment required for both methods tends to restrict the deployment of sensor-lines or cameras to a single place in a person's home. \RR{As a result, these methods can be used to measure gait velocity typically at one \emph{fixed} location and thus may miss important features of gait that occur in other locations in the home.}
%Both systems are also passive and thus require additional information to determine who generated the data when a walk is detected, although solutions have been proposed to address this issue~\cite{Austin2011}. 
In addition, camera based methods can also suffer issues with occlusion.  This can further limit the ability to measure gait in several common circumstances.  For example, if a resident moves their furniture or changes behavior (e.g., they regularly choose a path through the room for which the camera focal point has not been optimized). 
%Additionally, \ca{they are location aware, but currently too costly to deploy throughout a home. 
Lastly, depending on the use case, camera-based systems may be limited by a loss of privacy~\cite{demiris2009older,boise2013willingness}.

We demonstrate in this paper that transition times can accurately predict walking speed.  One advantage of this approach is that it is less sensitive to sensor placement.  Another advantage is that it utilizes common infrared sensors that are already deployed in homes for security purposes~\cite{Kaye2011,Cook2011,Rantz2013b}.  Furthermore, this approach provides location-specific gait velocity with less than \RR{2.5 cm/s} of error, on average. The prediction model is developed using sensor-line estimated gait velocity as ground truth for proof of concept. Once developed, the model can be used without having another gait measurement system simultaneously deployed. 

\begin{figure}
\centering
\includegraphics[width=0.6\linewidth]{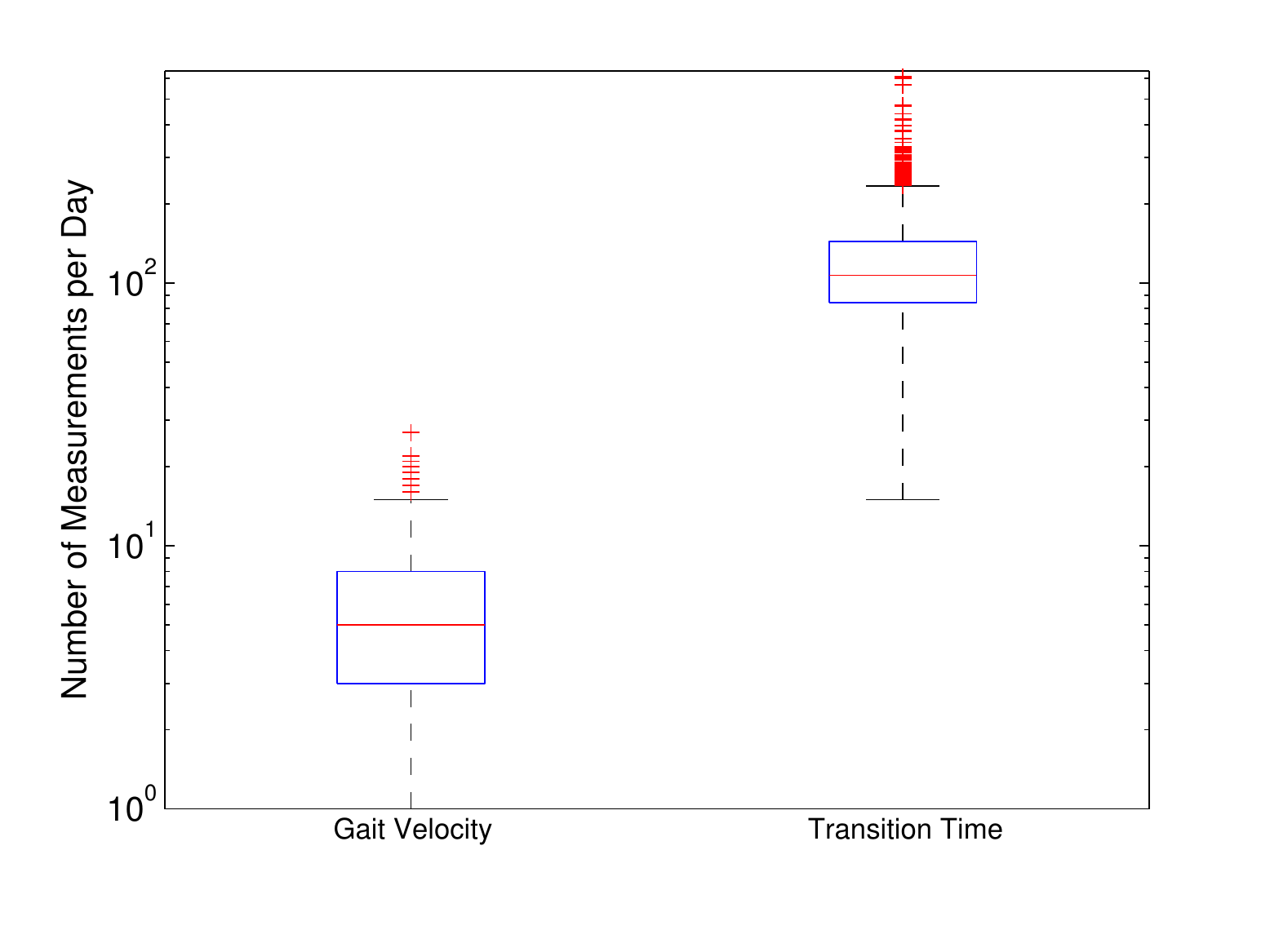}
\caption{Comparison of number of measurements per day: Gait Velocity versus Transition Time.}
\label{fig:statGAITvsTT}
\end{figure}
The proposed approach offers several improvements over the current state of the art.  First, since velocity is measured everytime a person moves between rooms in their home, the proposed method can gather 20-100x more estimates of gait velocity per day than other unobtrusive systems. \RR{In Fig.~\ref{fig:statGAITvsTT} we present the number of measurements for transition time and gait velocity for one representative subject. For this subject the average number of gait velocities measured using the walking line was approximately 6 per day, whereas the average number of velocities measured using transition time was approximately 121 per day.”} Our approach is also less sensitive to sensor placement. We demonstrated that room transition times can be used to accurately predict gait velocity than other approaches and it uses sensors that are usually already deployed in a smart home.  Our approach also provides location-specific gait velocity (e.g., the speed to or from the bathroom or phone) without incurring additional cost. This location-specific system will allow further investigation into the interplay between gait velocity and context, which may account for some observed variability in speed throughout the day.  Finally, studies have indicated that measuring fast vs. slow vs. average walking speeds~\cite{Fitzpatrick2007}  and measuring variability in these walking speeds~\cite{Ijmker2012} may be critical in passively assessing patient health using in-home monitoring.  The approach that we present using transition times enables the stable measurement of fast, slow, and average walking speeds throughout the home.   The fact that this method can acquire far more velocity estimates than a walking line located in a restricted location in the home could potentially enable earlier detection of movement-related health changes.

\begin{figure}
\centering
\includegraphics[width=0.8\linewidth]{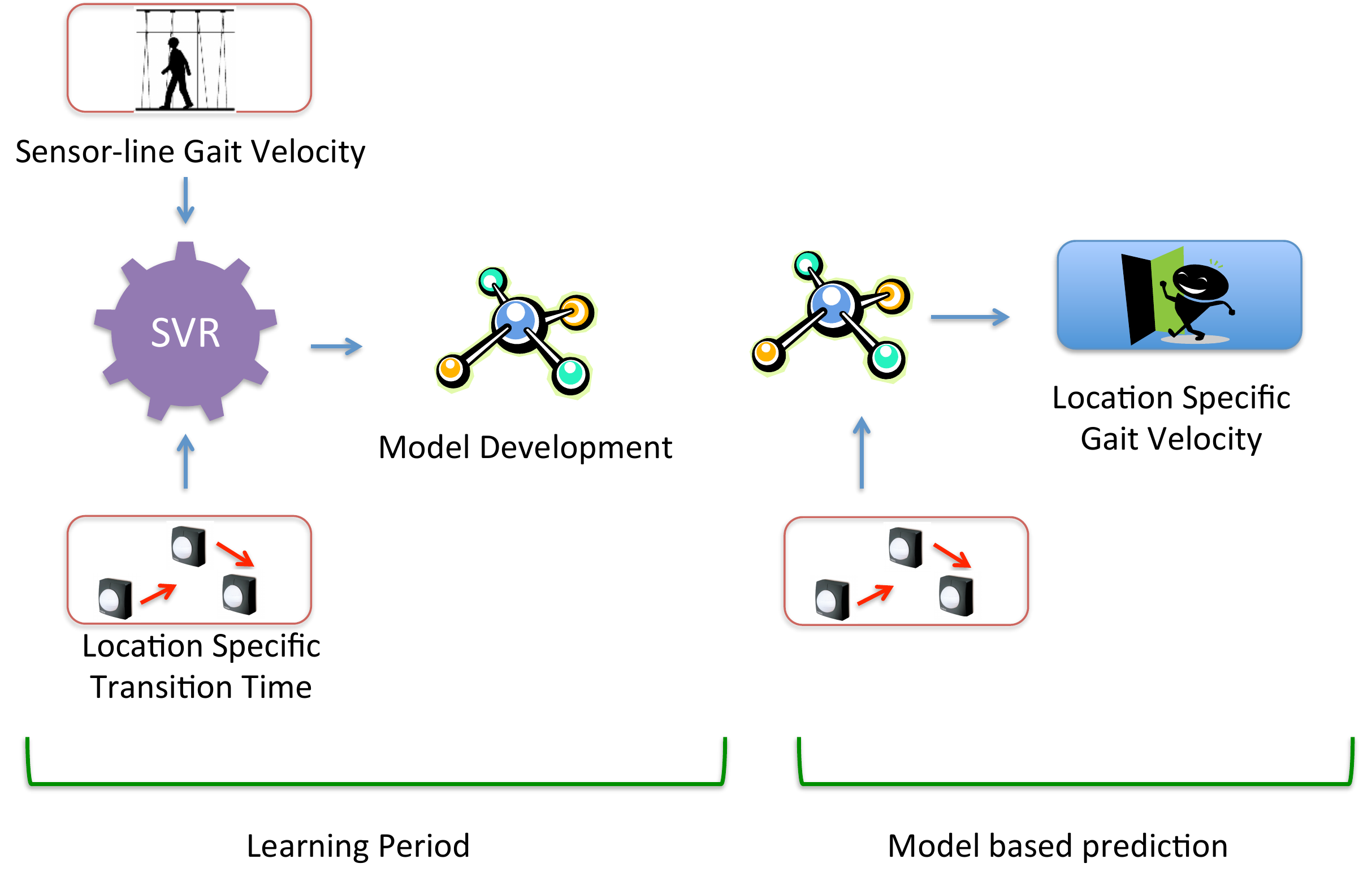}
\caption{Location-specific gait velocity prediction.}
\label{fig:architechture}
\end{figure}

%\section{Study Description and Feature Selection}
\section{Study Description }
%\subsection{Data}
In this study we used a data set collected from 74 non-demented older (mean age  85.9 years) men and women living independently who were part of the Intelligent Systems for Assessing Aging changes (ISAAC) project conducted by Oregon Center for Aging and Technology (ORCATECH) Living Laboratory. The ISAAC study is a longitudinal community cohort study using an unobtrusive home-based assessment platform installed in the homes of many seniors in the Portland, OR (USA) metropolitan area and is described in detail elsewhere~\cite{Kaye2011}.  \DA{Subjects living alone in the ISAAC cohort lived in a variety of different home sizes with 5.7 rooms on average (SD=2.2 rooms) and a range from studio style (3 rooms) to large houses (16 rooms).  For these homes, the average size is 900 ft$^{2}$ (SD=448 ft$^{2}$) with a range from 324 ft$^{2}$ to 3560 ft$^{2}$.  All subjects provided written informed consent and the study was approved by the Oregon Health \& Science University Institutional Review Board (OHSU IRB 2353).}

\begin{figure*}
\centering
\subfigure[]{
\includegraphics[type=pdf,ext=.pdf,read=.pdf,width = 1\linewidth]{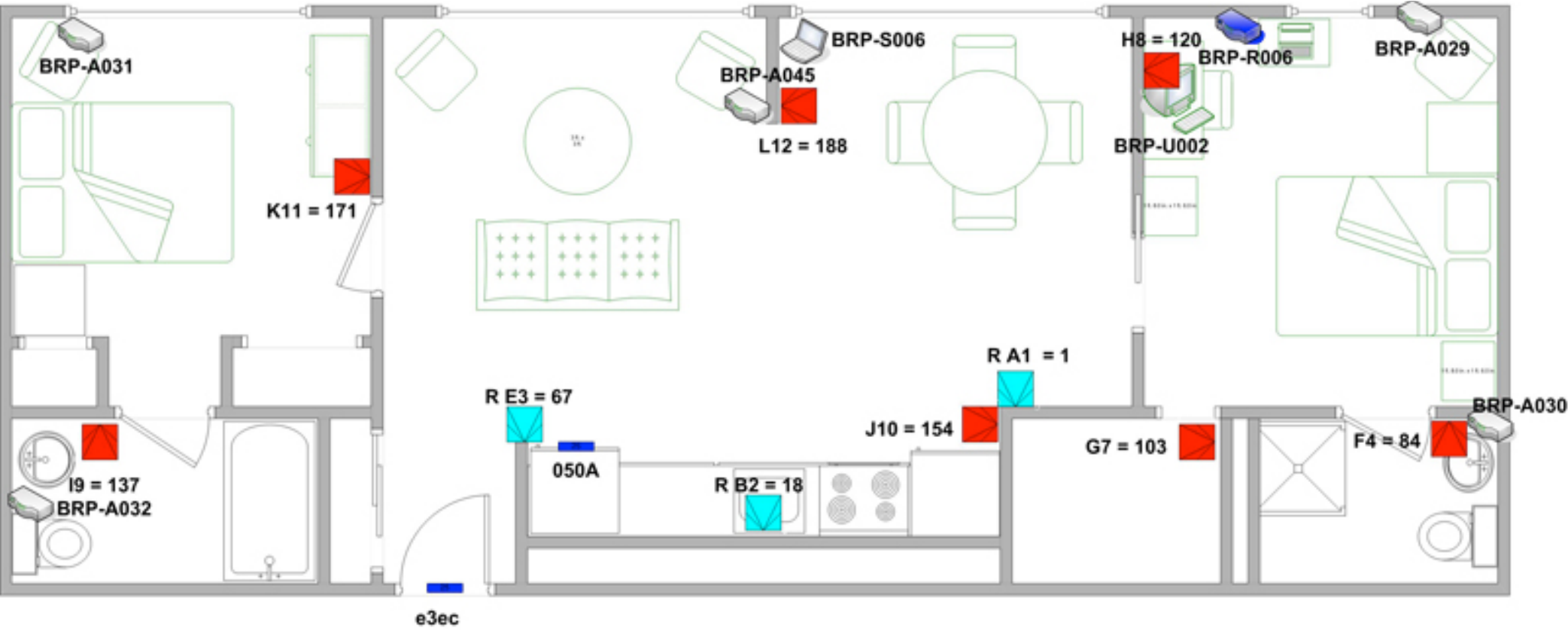}
\label{fig:home.13}
}
\subfigure[]{
\includegraphics[width = 0.3\linewidth]{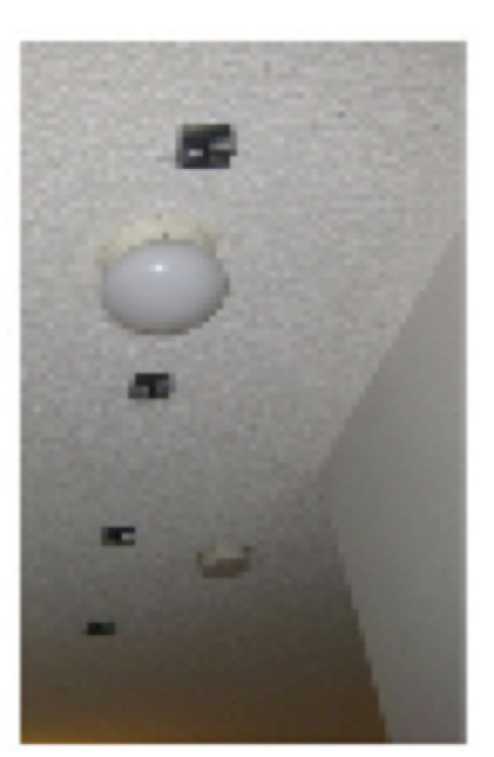}
\label{fig:ceilingImage.eps}
}
\subfigure[]{
\includegraphics[width = 0.3\linewidth]{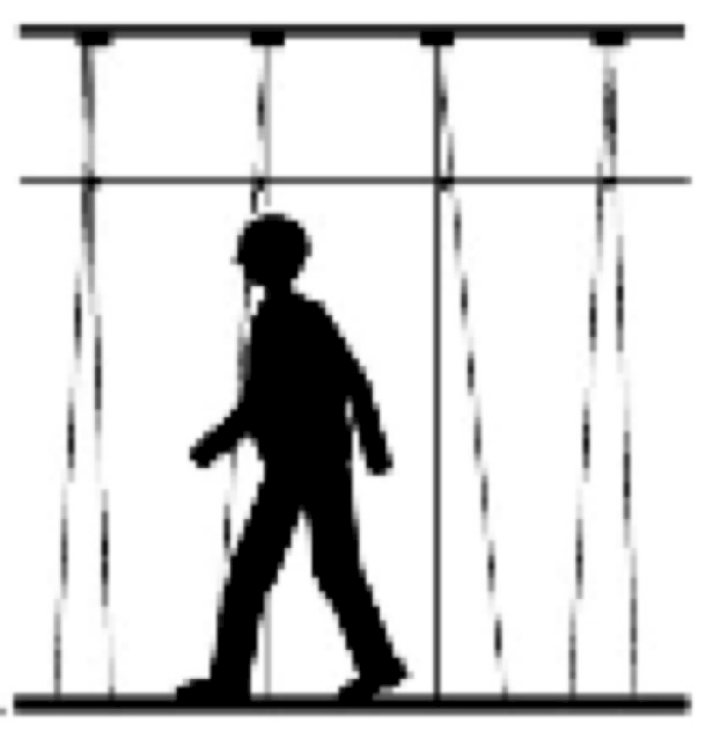}
\label{fig:walkingImage.eps}
}
\caption{(a) Sensor distribution in participant's home. Squares in red and cyan are representing passive infrared sensor. Blue rectangles are reed switches (b) sensor array in the ceiling (c) Schematics of a person walking though the sensor line.}
\label{fig:miscelleneous}
\end{figure*}

\subsection{Data Collection and Preprocessing}
\subsubsection{Data}
\DA{\RR{Two} types of data were used from each participant.  First are the \emph{transition times}, estimated as the time between sensor firings from X10 motion detectors (MS16A; X10.com) positioned in adjacent rooms. Each room in each home has one (unrestricted) motion detector (Fig.~\ref{fig:home.13} illustrates the sensor placement in one participant's house) placed such that the 30 degree by 90 degree field-of-view (FOV) of the sensor spans as closely as possible the room in which the sensor is installed.  Different homes have different room sizes and floor plans, therefore the distance between the FOV of sensors in adjacent rooms depends on characteristics of the specific home in which it is installed.  This is why the support vector regression (SVR) must be trained separately for each pair of sensors as discussed below; the actual distance between the FOV of sensors placed in adjacent rooms of a home is not known a priori. All motion sensors used in this study have a 6 second refractory period and are time stamped to the nearest millisecond.  
%Finally, clinical gait velocity data assessed by a clinician with a 30 ft timed walk were collected as part of a battery of clinical tests administered at baseline enrollment and then annually.  
%In this study, we used both the in-home gait velocity and clinical gait velocity measurements as ground truth for calibrating and verifying the proposed model that maps transition time to velocity.  
Participants self-reported via an online survey such events as when overnight visitors were present in the house, days in which technical staff visited the home, or times when sensors did not function properly (e.g. due to a dead battery for  example). Data from days with overnight guests, when staff visited the home, or with sensor outage were excluded in our analysis. \RR{In total we had $44,753$ days of data from our $74$ participants. \ca{On average we had 630.32$\pm$325.8 days of data from each participant.}}} 
%
%\subsection{In-home Estimated Gait Velocity}

The second type of data is in-home gait velocity estimated from a \emph{sensor-line}~\cite{hagler2010unobtrusive}.  This data is used as the ground-truth for the purpose of this study. The in-home gait velocities are estimated using 4 infrared sensors with restricted fields of view positioned in a linear array on the ceiling \DA{with 2 ft between each sensor (See Fig.~\ref{fig:ceilingImage.eps})}.  As a person moves underneath each sensor in the array, they fire in order (See Fig.~\ref{fig:walkingImage.eps}).  The time between the firing and the position of each sensor are used to estimate the gait velocity. There are various events that can cause variability in this method of estimating walking velocity.  First, a participant may not pass through the sensor at a near-constant velocity (e.g., they may pass part way through, stop, then continue walking). Second, an undetected issue with one or more sensors may cause the data to be corrupted in a way that influences velocity estimation.  As a result, the restricted-view sensors can yield poor estimates of gait velocity that manifest as outliers in the data set.

\begin{figure}[ht]
\centering
\subfigure[]{
\includegraphics[width = .45\columnwidth]{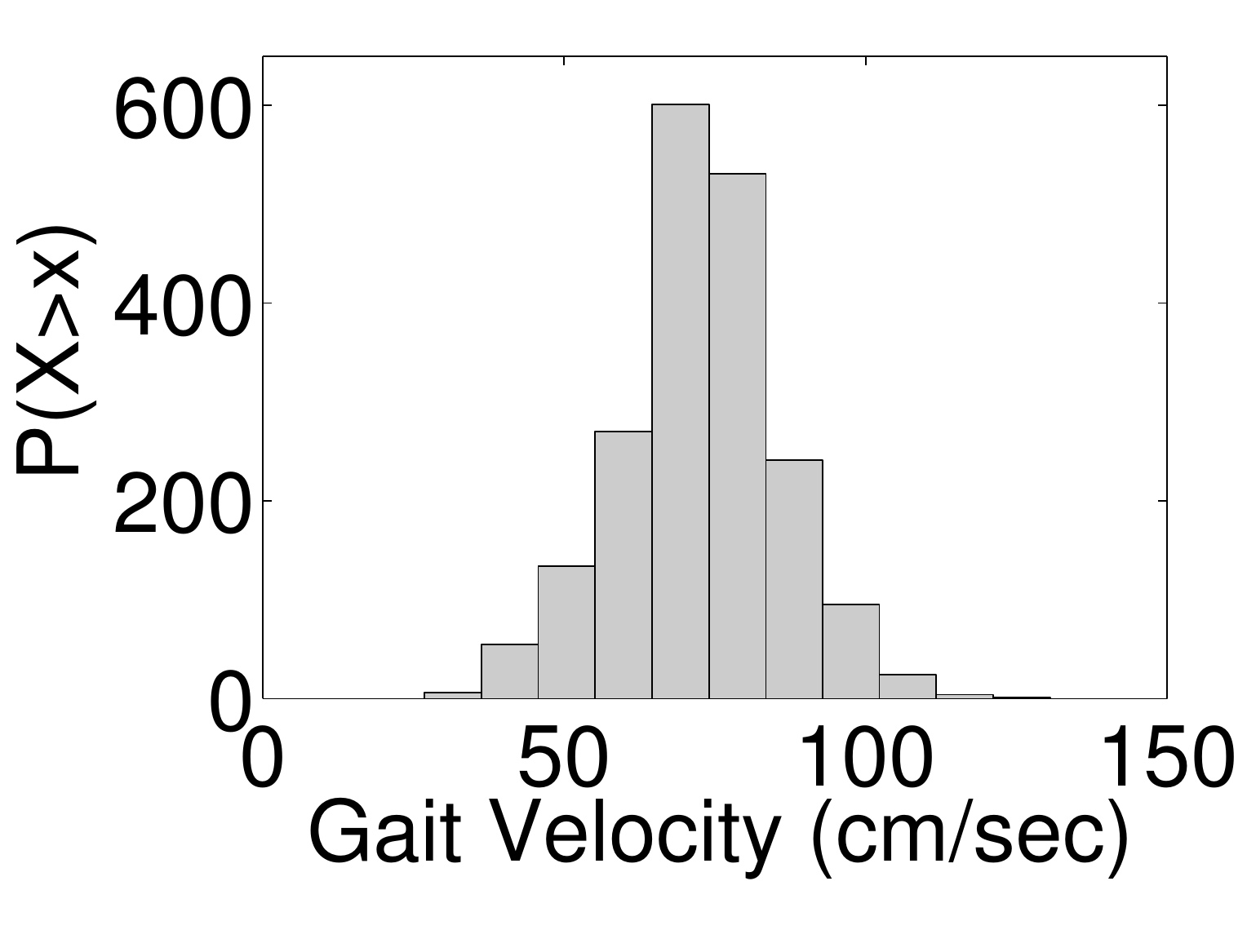}
\label{fig:histogram}
}
\subfigure[]{
\includegraphics[width = .45\columnwidth]{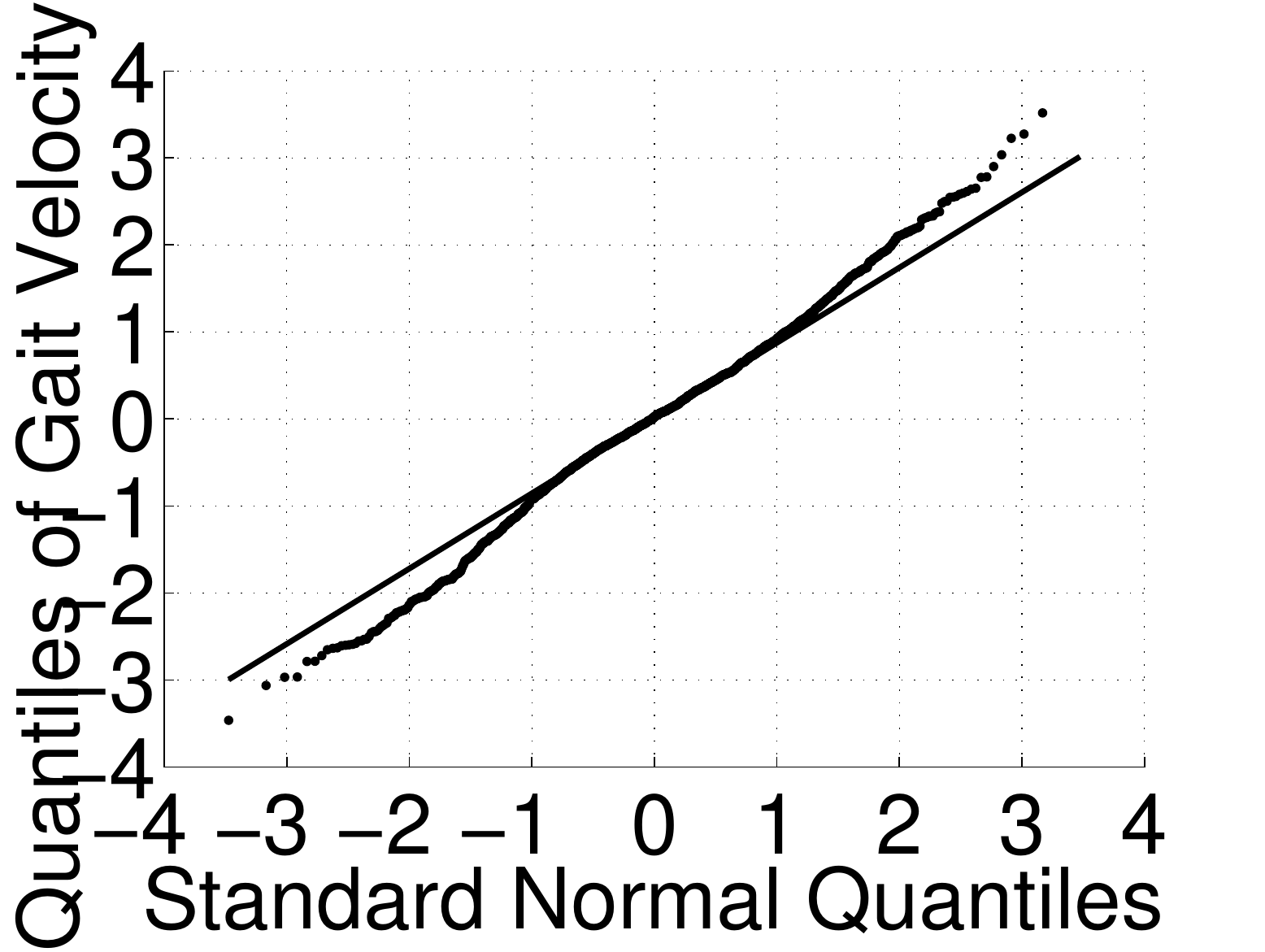}
\label{fig:QQplot}
}
\subfigure[ ]{
\includegraphics[width = 0.6\linewidth]{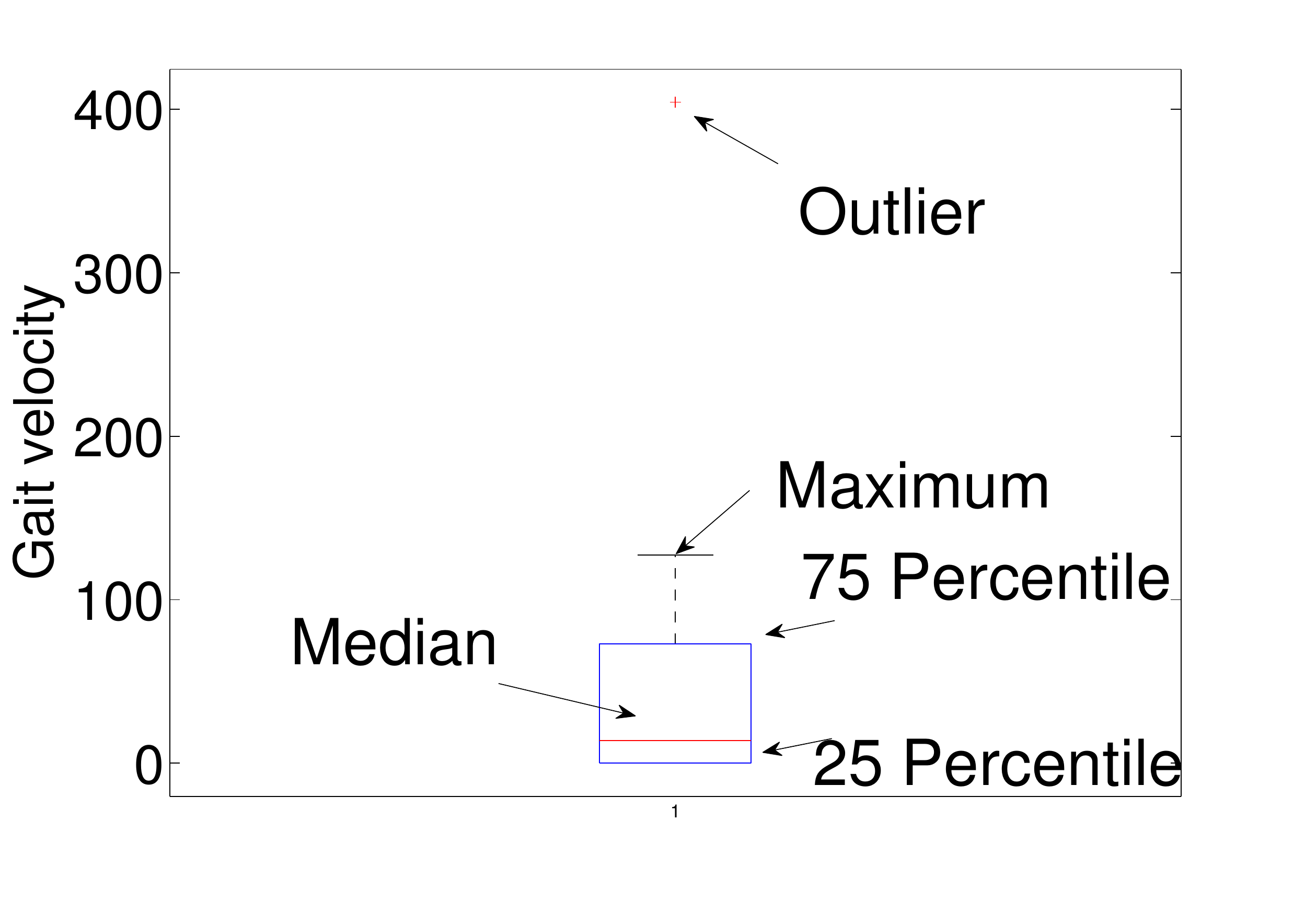}
\label{fig:boxPlot}
}
\caption{a) Distribution  of the gait velocity. b) Quantile-Quantile (QQ) Test of ``normal distribution'' of gait velocity estimated from restricted sensor line. c) Outlier Detection.}
\label{fig:walkSpedAnalysis}
\end{figure}

\subsubsection{Outlier Removal and Feature Selection}
\label{sec:DataAnalysisAndFeatureSelection}

In order to predict gait velocity from transition time, we first remove outliers, and then extract features from the transition times and ground truth gait velocities. Below we describe how data was processed using both the walking line and the transition times.

\noindent{\bf Gait-Velocity:}
\RR{In order to detect outliers we visually inspected the data and observed some values which are too large to be physiologically realizable gait velocities (e.g., 450 cm/sec etc; See Fig.~\ref{fig:boxPlot}). In order to exclude the outliers we only included data within two standard deviations of the mean velocity. After outlier removal, the majority of the values were approximately normally distributed with all measuerments less than the physiologically reasonable value of $150$ cm/sec (Fig.~\ref{fig:histogram}). After outlier removal the mean, standard deviation, minimum and maximum velocities were, 55.3 cm/s, 33.8 cm/s, 2.44 cm/s and 149.8 cm/s, respectively.}

There is not a one-to-one mapping between in-home gait velocities and transition times, since they are measured in different locations and with different sensors, thus do not co-occur. In other words we do not get a transition time and an in-home velocity that both correspond to the same movement at the same time.  Because of this, we aggregated the in-home velocity estimates across an entire day and used the mean gait velocity as the target to be estimated from the transition time. We use mean of gait velocity as the feature since  we conjecture that gait velocity is well represented by a normal distribution. 
%For normally distributed data, sample mean is identical to the mean of the population distribution. %We verified that the in-home velocity estimates approximately obey a normal distribution. For normally distributed data, the sample mean is representative of the true mean. Therefore, we could readily use the \emph{ mean} of the in-home gait velocity across an entire day as ground truth.  

We use Q-Q plots~\cite{wilk1968probability} ("Q" stands for quantile) to verify gait velocity is normally distributed. Q-Q plots is a graphical method for comparing two probability distributions by plotting their quantiles against each other. We plotted the quantiles of the empirical in-home gait velocity distribution with that of a standard normal distribution. If the two distributions being compared are similar, the points in the Q-Q plot will approximately lie on the line $y = x$. If the distributions are linearly related, the points in the Q-Q plot will approximately lie on a line, but not necessarily on the line $y = x$. In Fig.~\ref{fig:QQplot}, we plot the Q-Q test result for a representative participant. We found that the points in the Q-Q plot mostly lie on a line with a regression coefficient $r^2 = 0.9972$, and we observed this trend across all the participants. Therefore, the in-home estimated gait velocity is well approximated by a normal distribution. For normally distributed data, sample mean is identical to the mean of the population distribution. Therefore, we use the \emph{sample mean} as a representative feature for the in-home estimated gait velocity. %
%\subsubsection{Clinically Measured Gait Velocity}
%Clinically measured gait velocity was highly sparse as it is measured only once each year for each participant.  Due to the infrequency of the clinical assessment, we did not apply any aggregation on the clinically measured gait velocity. Instead, we used different time windows of the velocity estimates derived from room transition times centered around the clinical measurements to predict the clinically measured gait velocity.  We tested for time windows of 15 and 30 days and found no significant difference. We report the results for a window size of 30 days.
%\section{Gait Velocity Prediction using Transition Time}

\begin{figure}
\centering
\includegraphics[width = 0.8\linewidth]{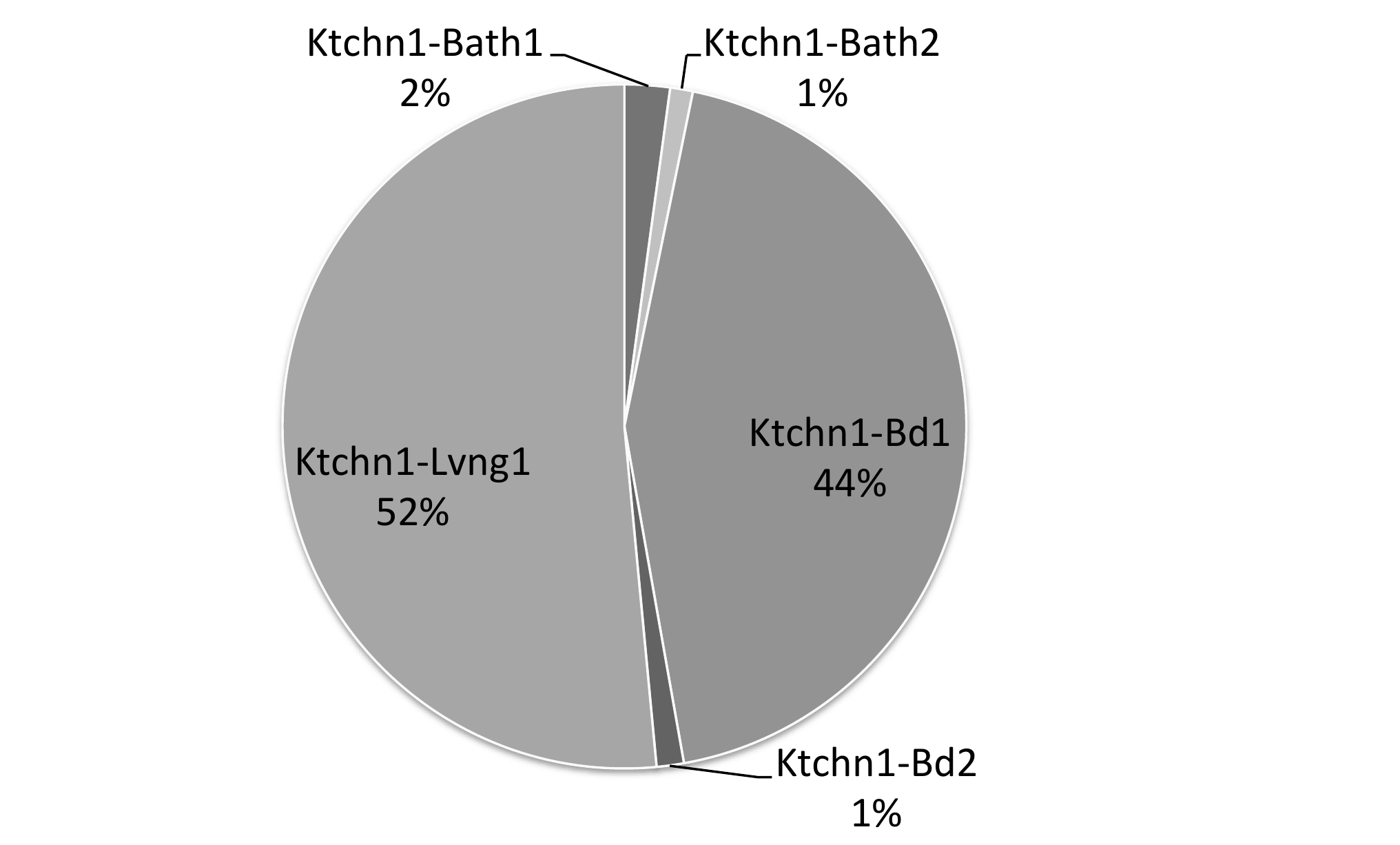}
\caption{Percentage transitions in various room pairs.}
\label{fig:roomVisitFrequency}
\end{figure}

\noindent{\bf Transition Time:}
%The transition times, after an outlier removal step, were used to predict both the in-home gait velocity and the clinical gait velocity.  
Some transitions between home locations do not occur frequently enough to permit a characterization of the distribution of the transition times.  This makes them unsuitable for use in velocity prediction.  This can occur, for example, when a room is infrequently visited such as a guest room. Note that the room transitions were mostly between adjacent rooms. However, because a sensor will occasionally not fire when a resident passes within the sensor’s field of view (due to the refractory period, for example), we will occasionally see instances where sensors in non-adjacent rooms are considered as transitions. To identify and remove these infrequent transitions, we removed all room pairs with $50$ or fewer observed transitions (\ca{over the entire period of data collection with the exception of when sensors were not functional or when visitors were reported to be in the home}) where $50$ was an empirically chosen threshold which allows reliable estimation of distributional parameters (e.g., the mean or different percentiles). We report the percentage of transitions for various room pairs in Fig.~\ref{fig:roomVisitFrequency} for one participant. Data taken from $921$ days were used to generate this plot. 
Some transitions, for example, kitchen to bathroom are very rare and should not be modeled due to the large statistical variability resulting from small sample sizes.

\begin{figure}[ht]
\centering
\subfigure[Kitchen-sensor line.]{
\includegraphics[width = 0.4\columnwidth]{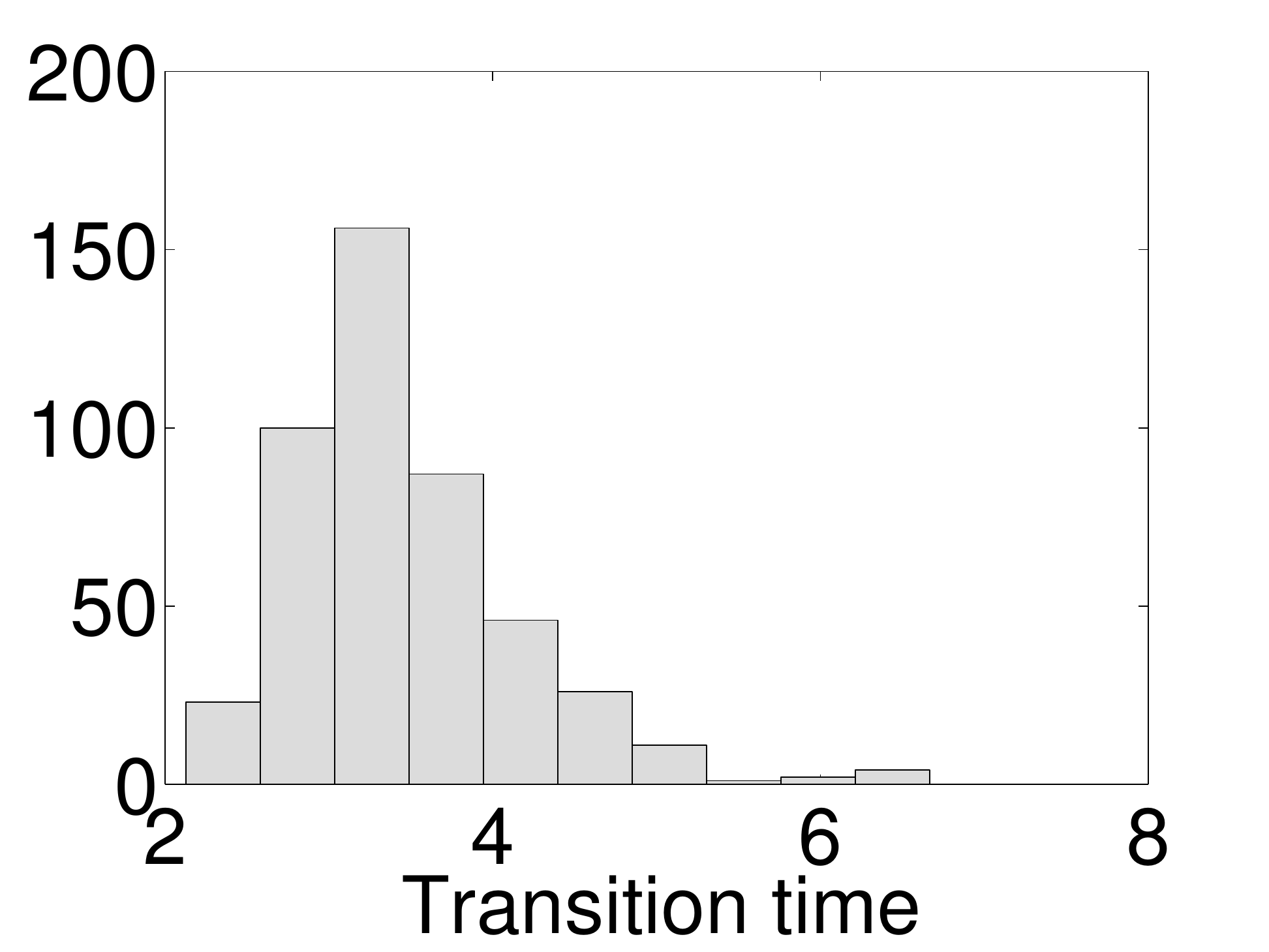}
\label{fig:svrFig1}
}
\subfigure[Living room-walk-in-closet.]{
\includegraphics[width = 0.4\columnwidth]{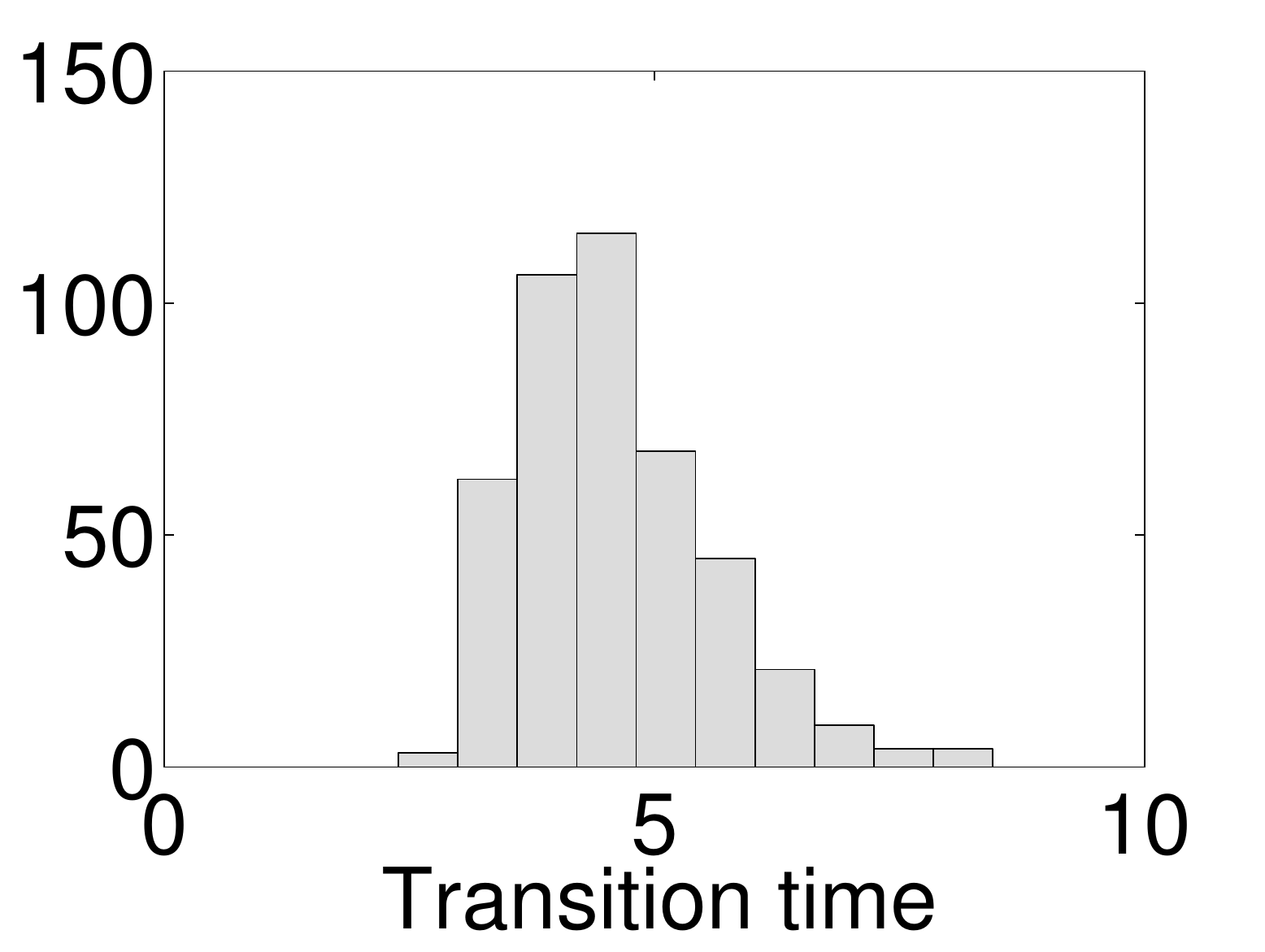}
\label{fig:svrFig3}
}
%\subfigure[Sensorline to bathroom 1.]{
%\includegraphics[width = 0.4\columnwidth]{images/skew4}
%\label{fig:svrFig3}
%}
\caption{Distribution (skewed) of transition time for two randomly chosen room pairs.}
\label{fig:skewedDistribution}
\end{figure}

The way transitions are measured can confound the speed at which a person travels between rooms and the time they spend in a room (dwell time).  This is because there is a refractory period of 6 seconds in the X10 motion sensors.  A person could, for example, trigger a motion sensor in the kitchen, wait 5 seconds before leaving the kitchen to go to the living room and then trigger the living room sensor.  Since the kitchen sensor could not fire again before the person left, the measured transition time would appear to be long when in fact it really represents a combination of dwell time and transition time.  For this reason the mean value of the transition time may not be the best feature describing movement speed (as was used for the in-home gait velocity). This is demonstrated in Fig.~\ref{fig:skewedDistribution}, where we observe that
% there is substantial variability in transition time
transition times are skewed.  Intuitively, smaller percentiles of the transition time distribution should be less likely to include dwell time. We therefore consider the \emph{$10th$, $15th$, and $20th$ percentile, along with the  first quartile, mean, and median }as potential features best summarizing the movement part of the transition time distribution.

\section{Support Vector Regression for Gait Velocity Prediction}
\label{subsec:svr}

Our approach to predicting gait velocity is based on learning the functional relationship between the transition times and gait velocity.  To learn this relationship, we used a support vector regression model, which is widely used for prediction~\cite{rana2013feasibility,rana2011adaptive}.

The complete description of SVR is outside the scope of this paper. However, we will provide intuition sufficient to understand the working principles of SVR. Consider a training set $\{(x_1,y_1),(x_2,y_2),...,(x_\ell,y_\ell)\}$, where $x_i$s are mean transition time and $y_i$s are mean gait velocity. Support vector regression computes the function $f(x)$ that has the largest $\epsilon$ deviation from the actual observed $y_i$ for the complete training set.

\begin{figure}[ht]
\centering
\subfigure[No relation between $x$ and $y$.]
{
\includegraphics[width = 0.25\linewidth]{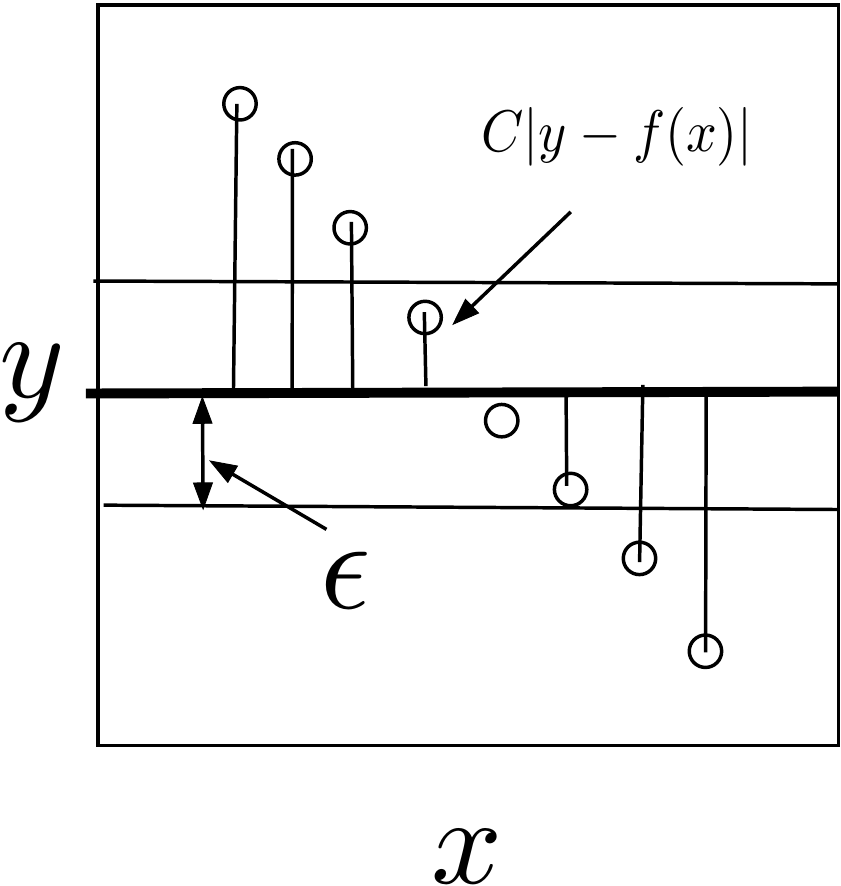}
\label{fig:svrFig2}
}
\subfigure[Linear regression.]{
\includegraphics[width = 0.25\linewidth]{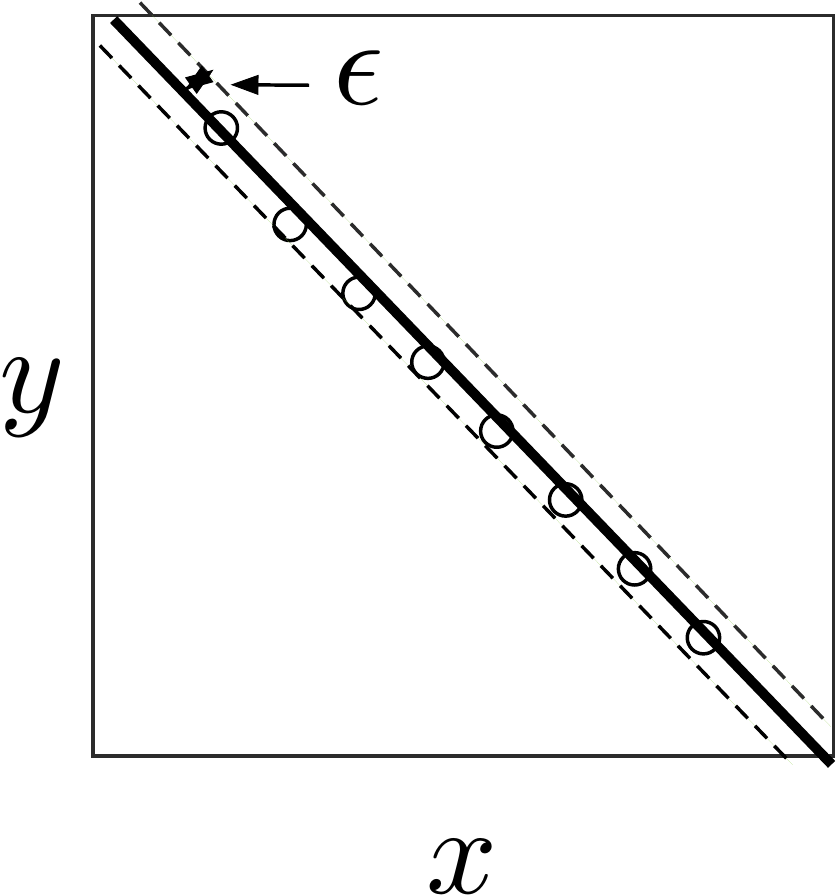}
\label{fig:svrFig1}
}
\subfigure[Linear SVR.]{
\includegraphics[width = 0.25\linewidth]{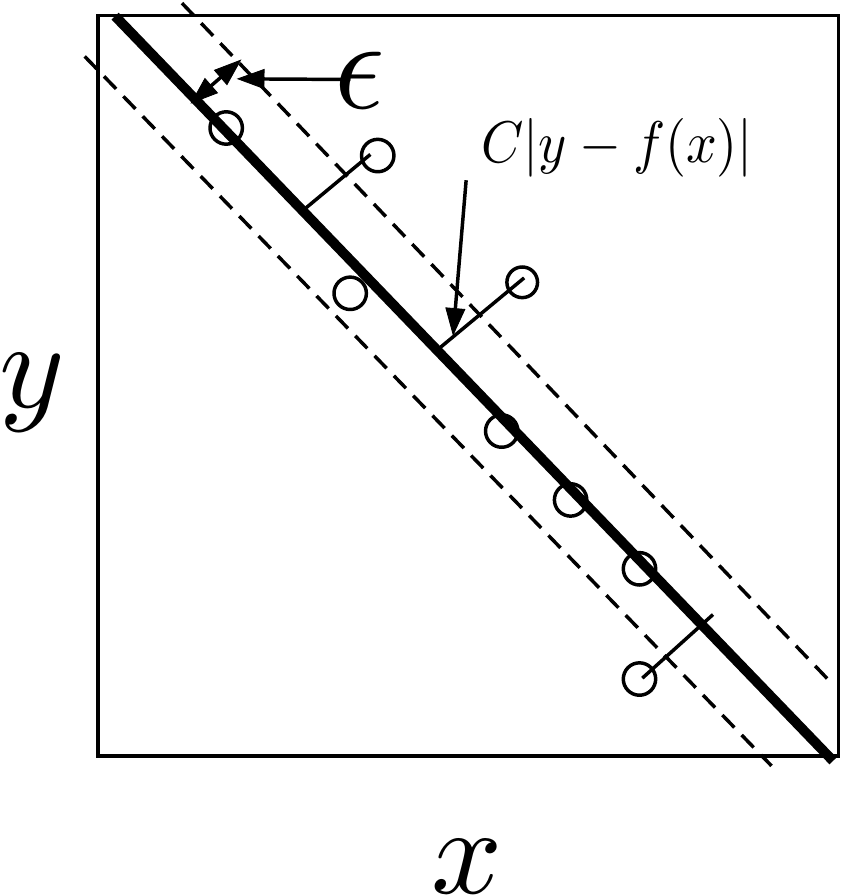}
\label{fig:svrFig3}
}
\caption{Support vector regression explained.}
\label{fig:svrExpalined}
\end{figure}

Let us assume the relationship between the variables is linear of the form $y = \omega x + b$, where $\omega$ (weight vector) and $b$ are parameters to be estimated. Fig.~\ref{fig:svrFig2} shows a few possible linear relationships between the points $x$ and $y$. The solid line in Fig.~\ref{fig:svrFig3} shows the SVR line given by $f(x) =  \omega x + b$. The cylindrical area between the dotted lines shows the region without regression error penalty. In the SVR literature this area is considered as the measure of complexity of the regression function used. Points lying outside the cylinder are penalized by an $\epsilon$-insensitive loss function~\eqref{eqn:lossFunction}~\cite{Vapnik:1995:NSL:211359} given by $|\xi|_\epsilon$.
\begin{eqnarray}
|\xi|_\epsilon := \begin{cases}
0 {\hspace{2 cm}		\mbox if } |\xi| \leq \epsilon\\
|\xi| - \epsilon			\mbox{ \hspace{1 cm}	 otherwise.}
\end{cases}
\label{eqn:lossFunction}
\end{eqnarray}
Now lets us explain the implication of a few different values of $\omega$. In the extreme case when $\omega = 0$ (as in Fig.~\ref{fig:svrFig2}), the functional relationship between $x$ and $y$ is least complex or in other words there is no relationship between $x$ and $y$. Therefore the overall error is very high. Next Fig.~\ref{fig:svrFig1} represents the case where the training data fits the solid line quite well. The solid line represents the classical regression analysis, where the loss function is measured as the squared estimation error. Note that although the solid line fits the data well, the cylindrical area between the dotted line is small, which means that the model will not generalize as well in predicting new data. SVR seeks to find a balance between the flatness of the area amongst the dotted lines and the number of training mistakes (see Fig.~\ref{fig:svrFig3}).

\begin{figure}
\centering
\includegraphics[width=0.7\linewidth]{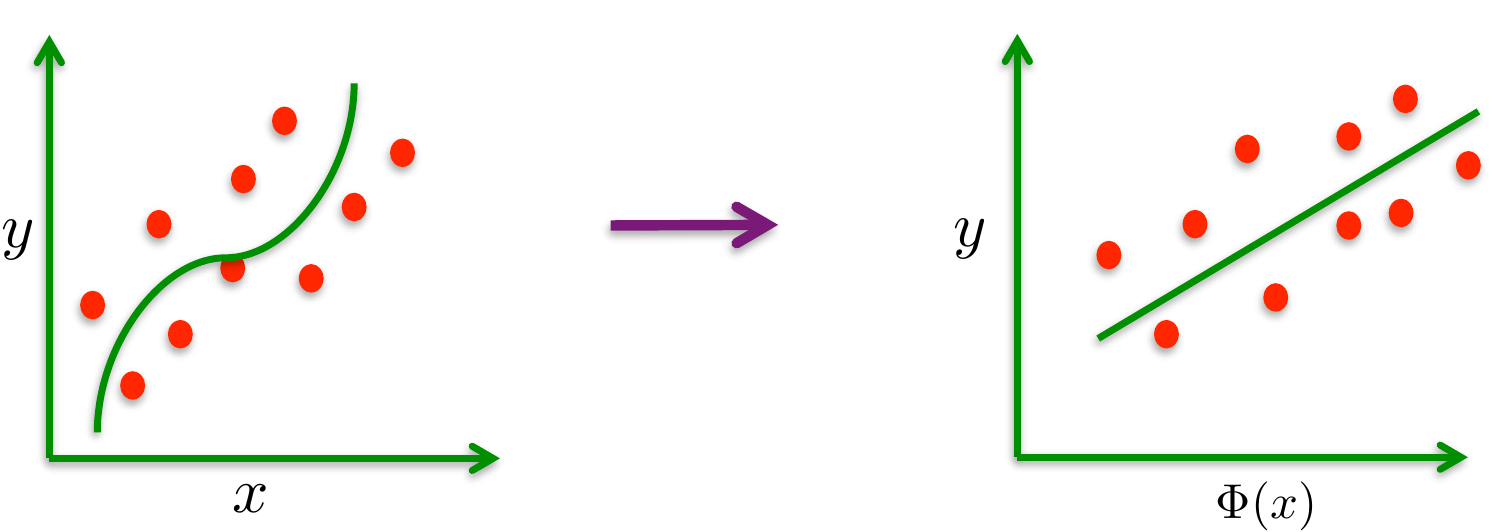}
\caption{Feature Space Transformation.}
\label{fig:transformSVR}
\end{figure}

\begin{figure}
\centering
\includegraphics[width=1\linewidth]{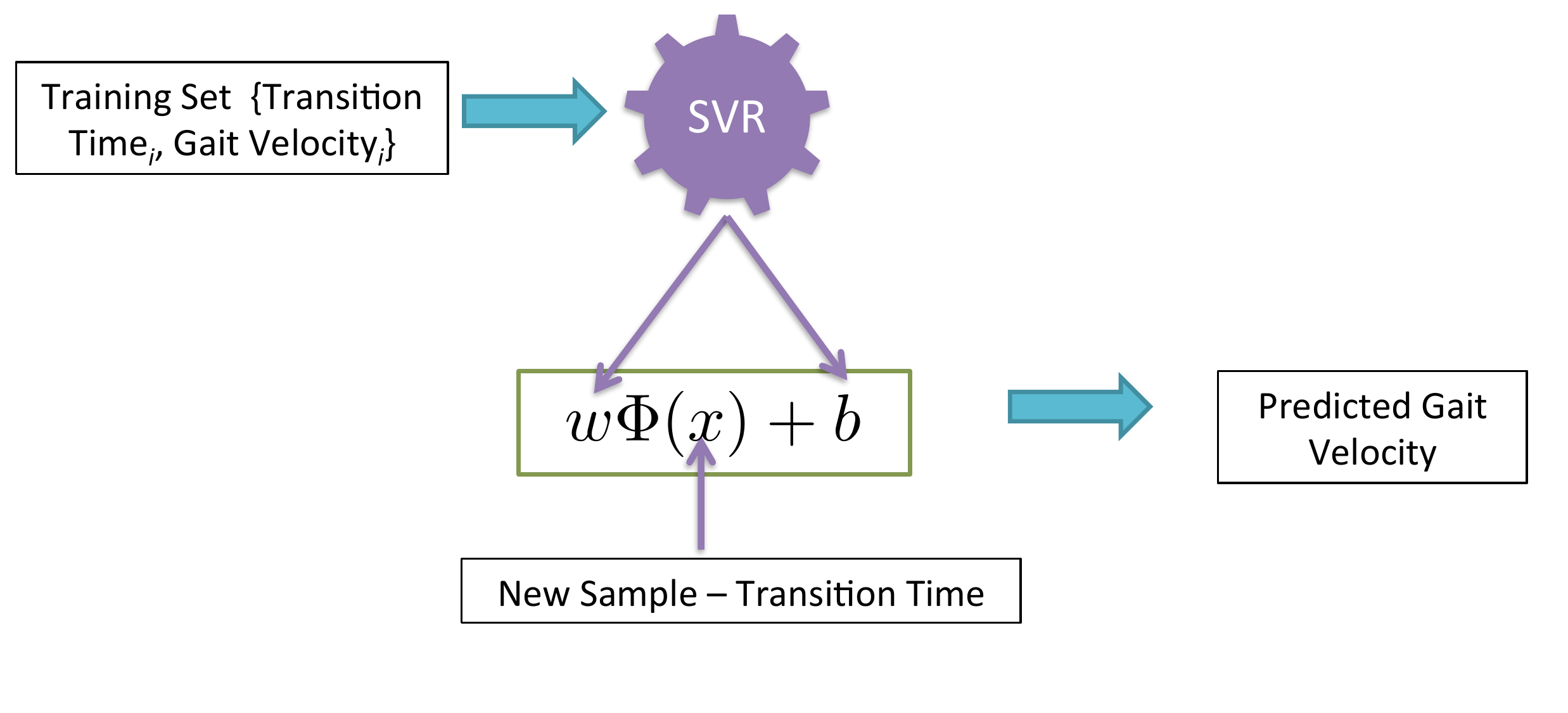}
\caption{Gait Velocity Prediction using SVR.}
\label{fig:finalFrameWork}
\end{figure}

Note that in many cases the relationship between the variables is non-linear as shown in the left diagram in Fig.~\ref{fig:transformSVR}. In those cases the SVR method needs to be extended, which is done by transforming $x_i$ into a feature space $\Phi(x_i)$. The feature space linearizes (right diagram in Fig.~\ref{fig:transformSVR}) the  relationship between $x_i$ and $y_i$, therefore, the linear approach can be used to find the regression solution. A mapping function or so called Kernel function is used to transform into feature space. There are four different functions which are frequently used as kernels within support vector regression: linear, RBF (Radial Basis Function), polynomial, and sigmoid. When the feature set is small, the RBF kernel is preferable over others. We use only one feature of transition time, therefore we use the RBF Kernel. However, we empirically verify that the RBF kernel performs better than the linear kernel. There are two parameters, namely $\gamma$ and $C$ (refer to~\cite{chang2011libsvm} for details) whose values need to be determined for best prediction. Here $C$ is the manually adjustable constant, and $\gamma$ is the kernel parameter which is formally defined as $K(x,y)= e^{-\gamma}||x-y||^2$. 
%The unknown parameters of the linear SVR $\omega$, $b$, and $\epsilon$ can be found as the unique solution of a dual of the primal problem  (see~\cite{smola1996regression}). There are a number of popular implementations of SVR in the literature including $\nu$-SVR~\cite{Schlkopf2000} and $\epsilon$-SVR~\cite{vapnik19196}. However, $\epsilon$-SVR or $\nu$-SVR just use different versions of the penalty parameter, the same optimization problem is solved in either case. We use $\epsilon$-SVR in our experiments. 
The overview of our SVR prediction framework is illustrated in Fig.~\ref{fig:finalFrameWork}.

\section{Results and Discussion}
\subsection{Simulation Setup}
%\subsubsection{Parameter and Model Selection for support vector regression}
We used the Matlab library LIBSVM~\cite{chang2011libsvm} to implement SVR. There are two functions: \texttt{svmtrain} and \texttt{svmpredict} for training and testing, respectively. We construct input feature $x^{'}$ using the transition time features discussed in Section~\ref{sec:DataAnalysisAndFeatureSelection}. The function \texttt{svmtrain} uses the input features to estimate $\omega$ and $b$. The prediction method \texttt{svmpredict} then uses these values and some other parameters (for details please review~\cite{chang2011libsvm}) to predict the gait velocity. We used five-fold cross validation to assess the model fits via RMS prediction error.  We input various features such as $10th$, $15th$ and $20th$ percentile, first quartile, mean and median of transition time to predict the gait velocity.
We report the transitions producing minimum prediction error while using different features in Table~\ref{tab:filter_coeff_2}. 

\subsection{Results}
%\subsection{Data Analysis}
We observe that the transition producing the minimum prediction error is variable. For example, the transition from Bathroom to Living is the best predictor for $10th$ percentile, whereas Kitchen to Refrigerator is the best predictor for $15th$ percentile. We even observe that this also varies person to person. For example, for participant $1$ the transition from Bathroom to Living is the best predictor for the $10th$-percentile, but this may not be true for participant $2$.  This is however realistic since transition times can be person or home - specific.
\begin{table}[t]
\small
\centering
\caption{Transitions producing the minimum prediction error.}
%\resizebox{!}{1.1cm}{
\setlength{\tabcolsep}{1pt}
\resizebox{6cm}{!} {
\begin{tabular}{|c|c|} \hline
\parbox{3.5cm}{\bf  Features (Giving minimum prediction error)}&{\bf Room pair}\\ \hline
$10th$ Percentile&Bathroom to Living  \\ \hline
$15th$ Percentile & Kitchen to Refrigerator \\ \hline
$20th$ Percentile &Bathroom to Living  \\ \hline
First Quartile&Kitchen to Refrigerator\\ \hline
Mean&Refrigerator to Kitchen\\ \hline
Median&Bed to Living\\ \hline
\end{tabular}
}
\label{tab:filter_coeff_2}
\end{table}

\begin{figure}
\centering
%\subfigure[]{
\includegraphics[width = 1\linewidth]{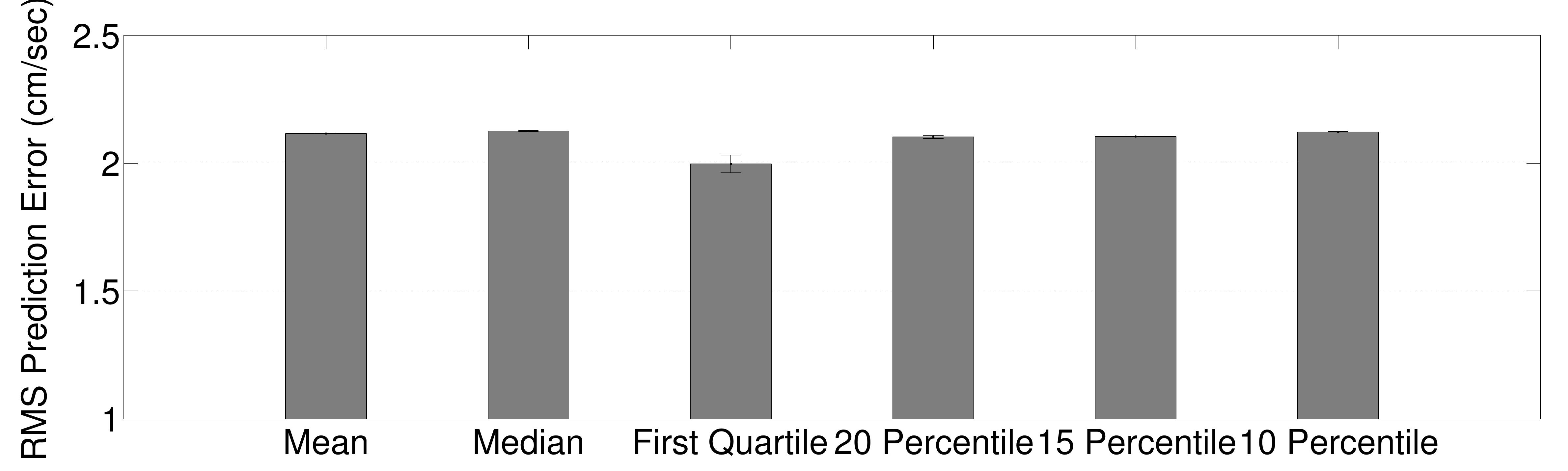}
%}
%\subfigure[]{
%\includegraphics[width = 1\linewidth]{images/clinicalNew.eps}
%\label{fig:clinicalWalkSpeedVersusTransitionTime}
%}
%\caption{In-home (a) and clinically assessed (b) gait velocity predicted from transition time across 74 participants. Various transition time features are used along x-axis. }
\caption{Gait velocity predicted from transition time across 74 participants. Various transition time features are used along x-axis. }
\label{fig:WalkSpeedVersusTransitionTime}
\end{figure}

In order to aggregate the prediction accuracy across all the participants, for each participant we calculate the prediction error for various features and repeat the calculation $100$ times. The mean and standard deviation of these prediction errors for various features are presented in Fig.~\ref{fig:WalkSpeedVersusTransitionTime}. 

Arranging the percentiles in decreasing order, we observe a minimum at the first quartile or $25th$ percentile. \DA{Quantitatively, the 25th percentile produces an average estimation error less than \RR{$2.5$} cm/s}. \RR{Intuitively, the transitions below the $25th$ percentile may not be typical; we speculate that these transitions may be observed when a person rushes from one room to the other room. Furthermore, the transitions above $25th$ percentile may be more likely to incorporate dwelling time. 
%Also note that the clinically assessed gait velocity is considered constant within the 30 days window. Therefore, from the prediction perspective there is less variability in the dependent variable, which offers the better performance for estimating clinically measured gait velocities.
}

\begin{figure}
\centering
\includegraphics[width=0.5\linewidth]{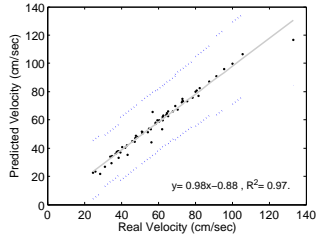}
\caption{Relationship between predictive and true gait velocity showing a highly linear
and strong correlation.}
\label{fig:compareTwoMean}
\end{figure}

\RR{Finally, we plot the mean ground truth gait velocity (measured using sensor-line) and the mean predicted (using $25th$ percentile of transition time) gait velocity for all the 74 participants in Fig.~\ref{fig:compareTwoMean}. Applying linear regression on the points we find that the points fit ($R^2=0.98$) a straight line with slope $1$ and intercept $-1.6$ with narrow $95\%$ confidence intervals. 
For perfect prediction all the points should be aligned to the line $y=x$ (slope $1$ and intercept 0). Therefore, our prediction performance is very close to ideal and approximately unbiased.}

\DA{Fig.~\ref{fig:compareTwoMean} also demonstrates the variability of the proposed estimator as a function of velocity.  The dotted gray lines in Fig.~\ref{fig:compareTwoMean} represents 95\% confidence intervals of the estimates, suggesting a reasonable spread of individual estimates around the average.  Additionally, the slight widening of the confidence intervals at the lowest and highest speeds indicate that these regions of the velocity curve are estimated less precisely, largely because there are fewer instances of the slowest and fastest walks with which to train the estimator.}

%We can 
%\RR{Lat but not the least let us explain how our proposed method offers \emph{location-specific} gait velocity estimation using Table~\ref{tab:filter_coeff_2}. From this table we observe that the transition from Bathroom to Living is the best predictor for $20th$ percentile. Therefore, counterintuitively we can use $20th$ percentile to estimate gait velocity between Bathroom and living room. We postulate that this will offer insight into the interplay of gait and context. }

%This can be

\subsection{Discussion}
\begin{table}
\centering
\caption{Comparison of various gait velocity estimation methods.}
\setlength{\tabcolsep}{1pt}
\resizebox{12cm}{!} {
\begin{tabular}{|l|l|c|c|c|}\hline
{\bf Method}&{\bf Accuracy}&{\bf Device-free}&{\bf Location/context} &{\bf Privacy} \\ 
&&&{\bf aware}& {\bf Conscious}\\ \hline
Room transitions &2.5 cm/s&Yes&Yes&Yes \\
(Current manuscript)& & & & \\ \hline
Accelerometer~\cite{Dalton2013} &4 cm/s compared to &No&No&Yes\\  
&GAITRite&&&\\ \hline
Sensor line~\cite{hagler2010unobtrusive}]&1.1 cm/s compared to &Yes&\parbox{3.5cm}{Yes(Limited to location where installed)}&Yes\\ 
&GAITRite&&&\\ \hline
Video~\cite{wang2013}& \% difference from&Yes&\parbox{3.5cm}{Yes (Limited to location where installed)}&No \\ 
&GAITRite (0.18\%)&&& \\ \hline
In-home gait mat~\cite{low2009initial}&Close to &Yes&\parbox{3.5cm}{Yes (Limited to location where installed)}&Yes\\ 
& GAITRite&&&\\ \hline
\end{tabular}
}
\label{tab:comparisonTable}
\end{table}

The results demonstrate that transition time can be used to accurately estimate gait velocity.  \ca{A comparison with other related technologies is presented in Table~\ref{tab:comparisonTable}. The other studies mentioned in Table~\ref{tab:comparisonTable} appear to have validated gait velocities against clinical gold standards but have not all reported results for the same accuracy measures, making a direct comparison of accuracy difficult. However, Table~\ref{tab:comparisonTable} clearly shows that our proposed method offers very high accuracy while being location-aware and non-invasive.}

%\DA{To give an example of how this accuracy is suitable for clinical populations, a recent study~\cite{Ries2009} demonstrated that individuals' variance in a demented patient group can be upwards of 29 (cm/s)$^2$, suggesting that the average error in estimation of the proposed method is smaller than the normal variability that can be observed cross sectionally.  When using the method longitudinally to detect trends, the average error is differenced out and thus trends can be estimated with increased accuracy.}

Our methodology has three main advantages over other in-home sensing based technologies. First, we can estimate a gait velocity every time a person switches rooms in their home.  This can produce substantially more estimates of gait velocity than are available from competing methodologies.  Using the sensor-line as an example, we typically measure between 0 - 20 gait velocities a day.  On the other hand, a resident can move between rooms from 200-500 times a day.  As a result, measuring gait speed through transitions can provide a very dense set of velocity measurements over the course of each day, which can improve estimates of aggregated parameters (e.g., mean or maximum daily walking speed) and provide a rich set of time-specific movement information.

Second, our system does not require expensive dedicated sensors, such as the camera based sensors. The cost reduction from not having dedicated sensors for gait velocity estimation can be considerable, especially when scaling to many homes.

Finally, or perhaps most importantly, because the proposed system can estimate a gait velocity every time someone switches rooms, the proposed method produces multiple estimates of different {\it location-specific} gait velocities throughout the course of the day.  This context rich data will allow future studies to explore and account for variability in gait velocity associated with location (e.g., hurrying to the bathroom or to answer the phone), which is not currently possible on a large scale.  Combining this with the time information (as mentioned above) will further allow the study of time-space velocity trajectories - the study of how fast and when people move through their environment.

There are also some shortcomings with the proposed methodology.  \ca{First, we focused on single resident homes.  Living alone is a relatively common scenario for our target population of older adults~\cite{greenberg2011profile}, but may limit generalizability to other populations.  However, our approach may be extended to dual or multiple occupancy homes in the future as recent studies have shown that passive motion sensors can be used for resident identification and disambiguation of passively collected data~\cite{Banerjee2012,Austin2011}. Besides multiple residents, pets can also potentially interfere with our proposed system. We have not conducted any experiments with pets but it is reported that the range of the motion detector can be altered to eliminate motions close to the floor such as from pets~\cite{caudle2004illumination}.}

\begin{figure}
\centering
\includegraphics[width=0.65\linewidth]{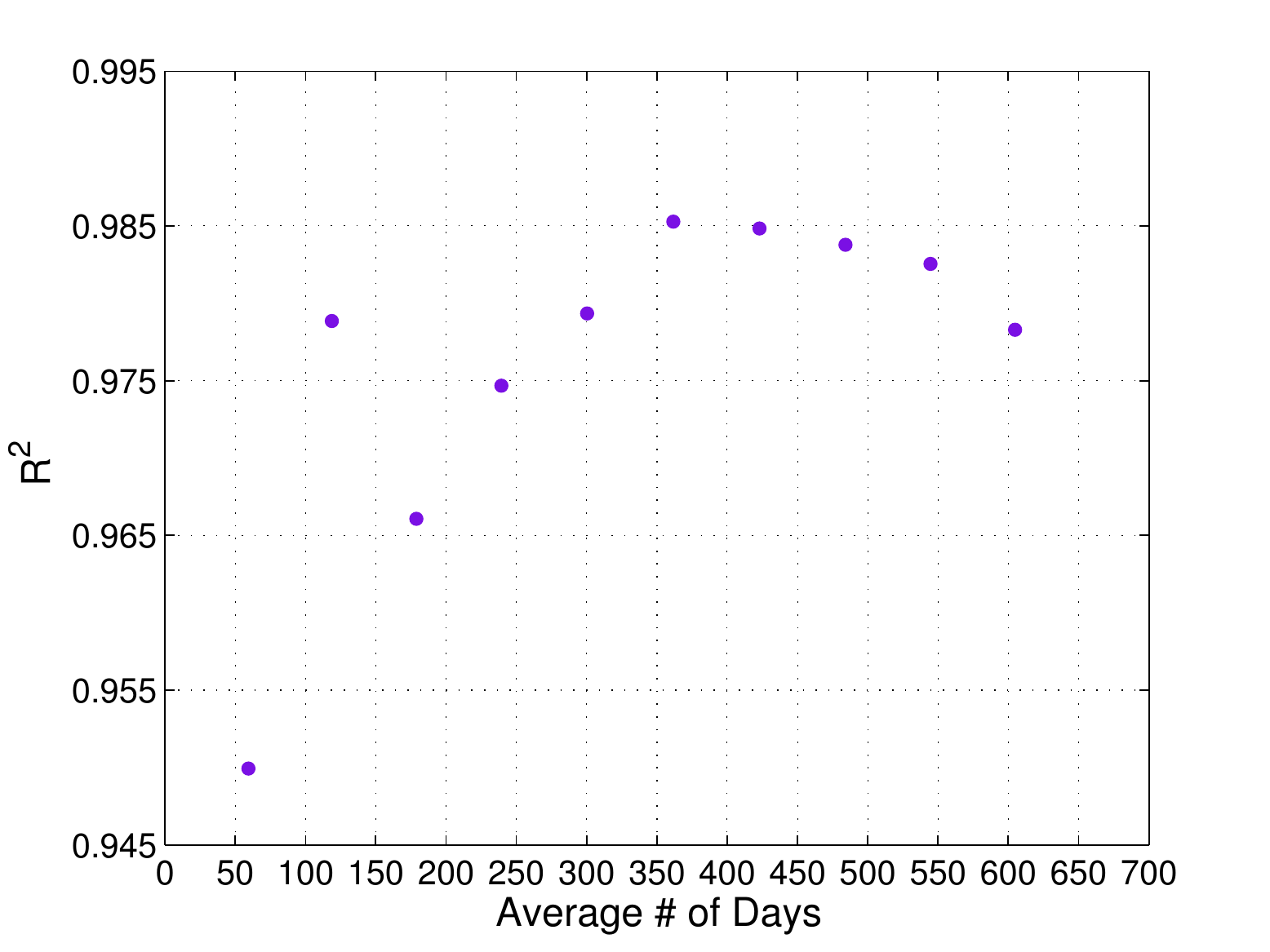}
\caption{Accuracy Versus Learning Period}
\label{fig:LearningTimeNew}
\end{figure}

Another shortcoming is that the proposed method requires that the model be trained using ground truth gait velocity collected within each residents' home.  In this study, a sensor-line was used (although a camera based or other method could also have been used). In practice, deploying a sensor-line increases the cost of the system (4 extra sensors) and maintenance expenses (battery changes, etc.). However, a sensor-line (or other ground truth system) is required for only a short period of time to train the system, so it could conceivably be removed and deployed in a different residence after the training period is complete.  In order to determine a sufficient training period, we generated Fig.~\ref{fig:LearningTimeNew}. In this figure we report the $R^2$ estimates (as in Fig.~\ref{fig:compareTwoMean}) for various training period. We observe that when the mean training period is around 250 days, the $R^2$ estimate is $0.97$, which resembles a good estimation of the gait velocity. 
We also observe that the $R^2$ estimate does not improve significantly when the mean training period is beyond 250 days. On average we have 630 days of data from Each participant. Therefore, we can construct an accurate model using less than 40\% of the training period. 
%we can produce a reasonably good estimation using less than 40\% of  
%The only a 40\% of data points \DA{RAJIB, a more useful metric would be average time needed to train the SVR.  Can you calculate and replace the proportion with the time in this section?  For example, how long is 40\%?} can offer similar accuracy as obtained when using 100\%\footnote{Note that we perform 5 fold cross validation to determine the prediction accuracy} of data.  
An alternative solution of the sensor-line produced gait-velocity training would be to use the transition times themselves as a proxy for true velocities.  This has been done previously to estimate mobility~\cite{Austin2014}.  In health monitoring applications, clinicians are often more interested in identifying change in gait velocity due to a health event, suggesting that absolute gait velocity may be less important than change.  For example, if a person has fallen, their gait velocity may change relative to the prior week.

Our approach works best (as with all in-home based systems) in populations who spend much of their time at home.  Becasue of this, a truly ubiqitous monitoring system would use our approach in conjuction with other in-home technology and ambulatory approaches such as GPS or wearable devices for truly pervasive and ubiqitous health monitoring. 

Finally, we are also keen to study the feasibility of sparse approximation methods~\cite{shen2013nonuniform,shen2011non,wei2012distributed,wei2013real,xu2011dynamic,east_ewsn} for gait velocity prediction. To this end sparse approximation methods have been used for classification\cite{wright2009robust,chew2012sparse}.  However, prediction can be formulated as a classification problem for future events. 

%\begin{figure}
%\centering
%\subfigure[]{
%\includegraphics[width=0.4\linewidth]{images/LearningTimeNew.eps}
%\label{AVLPa}
%}
%\subfigure[]{
%\includegraphics[width=0.4\linewidth]{images/statGAITvsTT.eps}
%\label{AVLPab}
%}
%\caption{a) Accuracy Versus Learning Period. b) }
%\label{fig:LearningTimeNew}
%\end{figure}

\section{Conclusion}
In this paper we demonstrated that room transition times can be used to accurately predict gait velocity.  Room transition times can be acquired using simple off-the-shelf IR sensors that are often deployed in home security systems or home-monitoring systems.  Using support vector regression for predicting gait velocity from the transition times, we show that the prediction accuracy of the approach is very high; quantitatively we can predict the gait velocity with \RR{less than 2.5 cm/sec error}.  This is demonstrated using data from $74$ participants collected over a five year period.  Using transition times to estimate gait velocity has several advantages over competing approaches such as: increased frequency of measurements, less sensitivity to sensor placement, and the ability to monitor time and location specific velocities.  In summary, the gait prediction approach described in this paper is simple, cost-effective, and highly accurate. It can be readily implemented in smart homes facilitating high resolution assessment of gait velocity, which has been shown to be an important predictor and indicator of healthy aging.

\section{References}
\bibliographystyle{plain}
\bibliography{referenceGaitTransitionTime,sigProc}

\end{document}